\documentclass[journal=jacsat,manuscript=article]{achemso}

\usepackage[version=3]{mhchem} 
\usepackage[colorlinks=true,linkcolor=red]{hyperref}

\author{Jiayan Xu}
\affiliation{Department of Chemistry, Princeton University, Princeton, New Jersey 08544, United States}
\author{Zheng Yu}
\affiliation{Department of Chemistry, Princeton University, Princeton, New Jersey 08544, United States}
\author{Abhirup Patra}
\affiliation{Shell International Exploration \& Production Inc., 200 N Dairy Ashford Rd, Houston, Texas 77079, United States}
\email{Abhirup.Patra@shell.com}
\author{Amar Deep Pathak}
\affiliation{Shell India Markets Pvt. Ltd. (Shell Projects \& Technology), Mahadeva Kodigehalli, Bengaluru, 562149, Karnataka, India}
\author{Sharan Shetty}
\affiliation{Shell India Markets Pvt. Ltd. (Shell Projects \& Technology), Mahadeva Kodigehalli, Bengaluru, 562149, Karnataka, India}
\author{Detlef Hohl}
\affiliation{Shell Information Technology International Inc., 3333 Highway 6 South, Houston, Texas 77082, United States}
\author{Roberto Car}
\affiliation{Department of Chemistry, Princeton University, Princeton, New Jersey 08544, United States}
\email{rcar@princeton.edu}

\title{
    Temperature-Induced Reorganization of Supported Zn$_\mathbf{3}$ Clusters on Cu(111): From Minimum-Energy Structures to Finite-Temperature Ensembles
}

\begin{document}


\newpage
\begin{abstract}
    Understanding the nature of catalytic active sites under reaction conditions remains a central challenge in heterogeneous catalysis.
    In the industrial copper/zinc oxide/alumina catalyst for methanol synthesis, small Zn-based species at the Cu interface have long been proposed as active-site candidates, yet their atomic-scale structure and stability remain controversial.
    Computational studies typically identify such species from optimized structures at 0 K, implicitly assuming that the minimum-energy configuration remains representative under reaction conditions.
    Here, we combine machine-learning-interatomic-potential-accelerated global optimization, molecular dynamics, and enhanced-sampling free-energy calculations to investigate the stability and dynamics of supported Zn$_3$(OH)$_3$ and Zn$_3$(OH)$_2$CHOO clusters on Cu(111)-based surfaces from 0 to 450 K.
    While compact triangular configurations are generally favored among the minimum-energy structures at 0 K, finite-temperature free-energy calculations reveal a pronounced shift toward extended linear configurations with increasing temperature.
    This configurational transition is driven primarily by entropic stabilization and cannot be inferred from potential energies alone.
    Molecular dynamics simulations further show that these clusters exhibit substantial mobility on pristine Cu(111) surface, beyond their configurational stability, the long-term persistence of these clusters depends on their surface mobility.
    Zn alloying in surface strongly suppresses diffusion, thereby stabilizing isolated interfacial Zn species.
    Together, these results demonstrate that the thermodynamically relevant structures of supported Zn-based clusters differ fundamentally from those predicted by static optimization and are governed by a competition between enthalpic and entropic effects.
    Our findings highlight the limitations of identifying catalytic active sites solely from 0 K structures and underscore the importance of explicit finite-temperature sampling in catalyst modeling.
\end{abstract}
\newpage

\section{Introduction}
Heterogeneous catalysis is traditionally understood through the concept of the catalytic active site, a specific atomic arrangement responsible for the adsorption, activation, and transformation of reactant molecules.\cite{Chorkendorff2017ConcceptsModernCatalysisKinetics}
Over the past decades, density functional theory (DFT) has become an indispensable tool for identifying such active sites and elucidating reaction mechanisms.\cite{Chen2020ChemRev}
In most computational studies, candidate active sites are generated from chemical intuition, surface thermodynamics, or global optimization techniques, and their relative stability is assessed primarily through comparisons of optimized total energies.
The resulting minimum-energy structures are then used as representative models for mechanistic investigations.
This approach has proven highly successful for many catalytic systems and has provided fundamental insights into structure–activity relationships.
However, the validity of describing catalytic active sites by a single minimum-energy structure becomes increasingly questionable when catalysts exhibit substantial structural flexibility under reaction conditions.\cite{Liu2005JACS,Zhang2020AccChemRes,Sun2021AccChemRes}
Experimental studies have shown that many heterogeneous catalysts undergo continuous restructuring, atom migration, changes in oxidation state, and dynamic interactions with adsorbates during operation.\cite{Shi2021JACSAu,Frey2022Science,Xu2023Science}
Under such conditions, multiple metastable configurations may coexist and interconvert on experimentally relevant timescales.
In these systems, the thermodynamically relevant active site is no longer necessarily the structure with the minimum potential energy at 0 K, but rather the ensemble of configurations populated at finite temperature.

Although finite-temperature effects can, in principle, be incorporated through ab initio thermodynamics,\cite{Reuter2001PRB} practical implementations often rely on a limited number of optimized structures and approximate entropic contributions using harmonic vibrational analysis.\cite{Christopher2004EssentialsComputationalChemistry}
Consequently, the thermodynamic ranking remains largely determined by differences in 0 K energies.
Global optimization methods substantially improve exploration of configurational space by identifying larger ensembles of low-energy structures,\cite{Grajciar2018ChemSocRev} but the resulting populations are still commonly inferred from optimized minima rather than directly sampled from the finite-temperature simulation.
As a result, the extent to which active sites can be reliably identified from local or global optimization alone remains an open question.
Addressing this challenge requires moving beyond static descriptions toward a statistical-mechanical framework in which active sites are defined by their finite-temperature populations and dynamic accessibility.
However, explicitly sampling the finite-temperature free-energy landscape with first-principles methods remains computationally prohibitive because it requires extensive molecular dynamics (MD) and enhanced-sampling simulations over long time and length scales.
Recent advances in machine learning interatomic potentials (MLIPs) provide a promising route to overcome this limitation.\cite{Xu2021PCCP,Cheng2024PrecisChem}
Trained on first-principles data, MLIPs can achieve near-DFT accuracy while accelerating MD simulations by several orders of magnitude, enabling explicit sampling of anharmonic free-energy landscapes and large-amplitude structural fluctuations that are inaccessible to conventional first-principles simulations.

Copper/zinc oxide/alumina (CZA) catalysts, the industrial benchmark for methanol synthesis from syngas and CO$_2$ hydrogenation,\cite{Ye2025Science} provide an ideal platform for examining those issues in catalyst modeling.
Despite decades of investigation, the atomic-scale identity of the active site remains actively debated.
In particular, the structure and chemical state of zinc species at the Cu interface under reaction conditions have been proposed to vary from isolated Zn atoms and CuZn alloys\cite{Behrens2012Science,Shi2022JACS} to partially oxidized Zn clusters.\cite{Kattel2017Science}
Experimental and theoretical studies further suggest that small Zn-based clusters may decorate Cu surfaces and participate directly in catalytic turnover.\cite{Kattel2017Science,Reichenbach2019JPCC,Koitaya2019ACSCatal,Jensen2024NatCommun}
However, the structure, stability, and thermodynamic relevance of these interfacial species remain poorly understood.
Because Zn-based clusters are expected to exhibit significant structural flexibility and multiple competing configurations, they represent an excellent model system for assessing the limitations of conventional 0 K structure identification strategies.

In this work, we combine MLIP-accelerated global optimization, enhanced sampling, and unbiased MD simulations to investigate Zn$_3$(OH)$_3$ and Zn$_3$(OH)$_2$CHOO clusters supported on Cu(111) surfaces relevant to methanol synthesis.
These clusters share a three-atom Zn core, which provides the simplest multinuclear model capable of exhibiting transitions between distinct geometries and thus serves as a prototypical system for investigating temperature-dependent structural behavior.\cite{Kumari2023JACS,Yang2025JACS}
We first establish the 0 K stability landscape by identifying low-energy triangular and linear Zn$_3$ structures on pristine and Zn-modified Cu surfaces.
We then determine their finite-temperature configurational stability by constructing free-energy profiles at 150, 300, and 450 K.
These calculations reveal a pronounced temperature-induced shift in the preferred cluster geometry: although compact triangular configurations are generally favored among the minimum-energy optimized structures, increasing temperature progressively stabilizes extended linear configurations through favorable entropic contributions, leading to a redistribution of the equilibrium population.
Beyond configurational stability, we further investigate the spatial stability of supported Zn$_3$ clusters using unbiased MD simulations.
Spatial stability describes the ability of supported clusters to remain isolated on the catalyst surface, rather than migrating and coalescing through sintering.
While small Zn$_3$ clusters exhibit substantial mobility on pristine Cu(111) surface, Zn alloying significantly suppresses diffusion of isolated interfacial Zn species, suggesting a reduced propensity for cluster migration and subsequent sintering.
Together, these results demonstrate that the stability of Zn$_3$ active-site candidates is governed by two complementary factors: configurational stability arising from the free-energy competition between triangular and linear ensembles, and spatial stability, determined by the resistance of supported clusters to surface diffusion and sintering.
More broadly, our work underscores the importance of explicit finite-temperature sampling for identifying catalytically relevant active sites.

\newpage
\section{Results and Discussion}
\paragraph{MLIP Development and Supported Zn$\mathbf{_3}$ Cluster Models.}
To enable extensive structure exploration and long-time MD simulations for CZA catalyst, we employed an active learning workflow developed in our previous work\cite{Xu2025ACSNano} to train a deep potential (DP) model for the Zn$\text{-}$Cu$\text{-}$O$\text{-}$C$\text{-}$H system, combining global optimization and MD to enable comprehensive sampling of the relevant chemical space.
The reference energies and forces were calculated at the PBE+D3 level, with PBE\cite{Perdew1996PRL} selected because it is widely used for Cu--Zn systems\cite{Reichenbach2019JPCC,Weinreich2020JPCC,Weinreich2021JPCC} and D3\cite{Grimme2010JCP} included to better describe dispersion contributions to cluster--surface and adsorbate--cluster interactions.
The model architecture and the training dataset are described in details in the \textbf{Methods} section.
The model reaches root-mean-squared errors of 0.005 eV/atom for energies and 0.056 eV/Å for forces with parity plots shown in \textbf{Figure S1}.
The RMSEs for energies and forces across different subsystems are summarized in \textbf{Table S1}.

To probe the stability and structural flexibility of supported Zn-based species, we selected Zn$_3$(OH)$_3$ and Zn$_3$(OH)$_2$CHOO as representative hydroxylated and formate-containing cluster models.
Hydroxylation is introduced to reflect the reducing and hydrogen-rich environment relevant to methanol synthesis conditions.
In addition, Zn$_3$(OH)$_2$CHOO is considered to explicitly incorporate formate, a key intermediate in CO$_2$ hydrogenation,\cite{Wu2017ACSCatal,Tameh2018JPCC} thereby enabling assessment of adsorbate-induced restructuring effects.
The choice of substrate models aims to investigate the role of Zn alloying in modulating the stability and mobility of supported clusters.
We adopted three Cu(111)-based surfaces with varying Zn concentrations in the top layer to represent different degrees of alloying and Zn incorporation at the interface (\textbf{Figure S2}).
Each surface adopts a four-layered $p(8 \times 4\sqrt{3})$ slab, which is sufficiently large to accommodate the Zn$_3$ clusters without significant interactions between periodic images.
The pristine Cu(111) surface represents the baseline metallic surface.
The single-Zn-site surface has one single Zn substitution in the top layer models dilute alloying, reflecting the initial stages of Zn incorporation.
Finally, the alloyed Cu$_3$Zn surface with 25\% Zn in the top layer approximates more extensive Zn$\text{-}$Cu mixing.
As the top surface layer contains 64 metal atoms, the corresponding surface Zn concentrations are 0\% ($0/64$), 1.5625\% ($1/64$), and 25\% ($16/64$), respectively.
These three surfaces span a realistic range of interfacial environments encountered in working CZA catalysts.
This systematic selection allows us to isolate how adsorbate coordination and substrate alloying collectively determine the stability and mobility of supported Zn-based clusters.

In the following, we first establish the configurational stability of Zn$_3$(OH)$_3$ on pristine Cu(111) surface at zero and finite temperatures, and then examine how surface Zn alloying and the formate intermediate modify the corresponding free-energy landscape.

\paragraph{Zero-Temperature Total-Energy Landscape.}
To establish the zero-temperature total-energy landscape, we first compared the low-energy structures of Zn$_3$(OH)$_3$ and Zn$_3$(OH)$_2$CHOO supported on pristine Cu(111) surface, single-Zn-site surface, and alloyed Cu$_3$Zn surface.
The candidate structures were generated via genetic-algorithm-based global optimization using the DP model for each cluster–surface combination, i.e. Zn$_3$(OH)$_3$ and Zn$_3$(OH)$_2$CHOO on the three surfaces, followed by DFT relaxation to validate the relative energies of the key configurations.
Technical details of the global optimization are provided in the \textbf{Methods} section.
The low-energy structures of both Zn$_3$(OH)$_3$ and Zn$_3$(OH)$_2$CHOO can be broadly classified into two distinct configurations (\textbf{Figure~S3}): a compact triangular configuration, in which the three Zn atoms form a planar triangle, and an extended linear configuration, where the Zn atoms adopt a more open arrangement, with one adsorbate (hydroxyl or formate) bridging the cluster and the surface.
Comparison with DFT calculations (\textbf{Figure~S4}) confirms that the DP model accurately reproduces the relative energetic ordering of these low-energy configurations across the different substrate compositions.

\begin{figure}
    \includegraphics[width=1.00\textwidth]{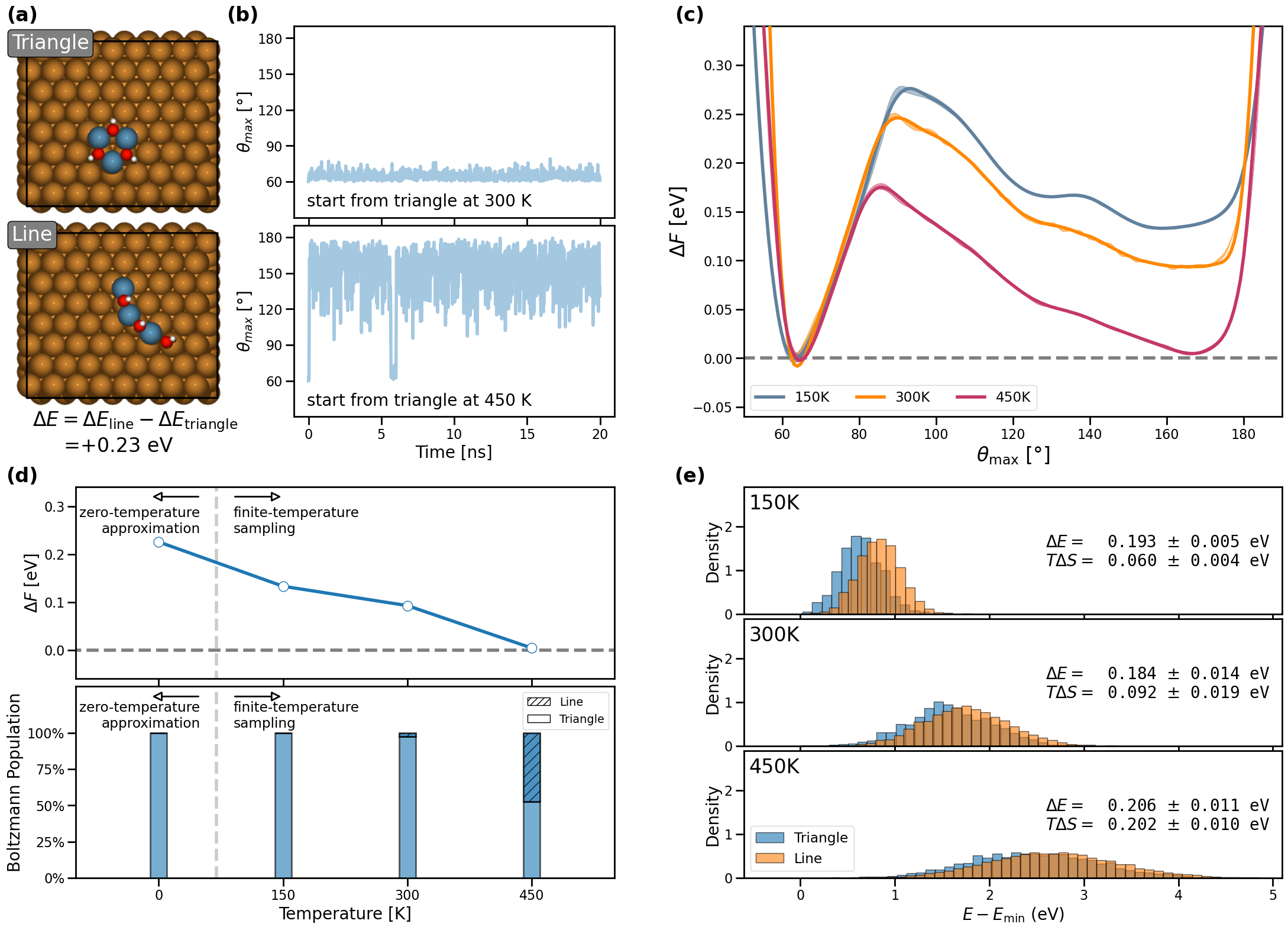}
    \caption{
        The configurational stability of Zn$_3$(OH)$_3$ on pristine Cu(111) surface.
        (a) Optimized triangular and linear configurations with their zero-temperature energy difference, $\Delta E = E_\mathrm{line} - E_\mathrm{triangle}$. Atom colors in (a): Cu, brown; Zn in the cluster, blue; O, red; H, white.
        (b) Representative unbiased MD trajectories at 300 and 450 K showing the evolution of the maximum Zn-Zn-Zn angle ($\theta_{\max}$), initialized from the triangular configuration.
        (c) One-dimensional free-energy profiles projected onto $\theta_{\max}$ at 150, 300, and 450 K, obtained from OPES simulations.
        The free energy is referenced to the triangular minimum.
        (d) Free-energy difference between the linear and triangular configurations ($\Delta F = F_\mathrm{line}-F_\mathrm{triangle}$) as a function of temperature (top), together with the corresponding Boltzmann populations (bottom).
        The zero-temperature energy difference is included for comparison.
        (e) Probability density distributions of the potential energy relative to the minimum sampled energy, $E-E_{\min}$, for the triangular (blue) and linear (orange) configurations at 150, 300, and 450 K.
        The corresponding potential-energy difference ($\Delta E = E_\mathrm{line}-E_\mathrm{triangle}$) and entropic contribution ($T\Delta S=\Delta E-\Delta F$) are listed in each panel.
        Shaded regions in (c), error bars in (d), and the uncertainties reported in (e) denote one standard deviation across cumulative reweighted estimates evaluated at 5 ns intervals. Where not visible, the error bars are smaller than the symbols.
    }
    \label{fig:zn3x}
\end{figure}

Although zero-temperature calculations identify the triangular configuration as the minimum-energy structure of Zn$_3$(OH)$_3$ on pristine Cu(111), catalytic reactions occur under finite-temperature conditions where thermal fluctuations may substantially modify the relative stability of competing configurations.
To establish how the active-site ensemble evolves with temperature, we first examine Zn$_3$(OH)$_3$ on pristine Cu(111) as a representative system.
\textbf{Figure~\ref{fig:zn3x}a} compares the optimized triangular and linear configurations obtained from DP-based global optimization.
The triangular structure is lower in energy by 0.23 eV than the linear structure, indicating that a static zero-temperature description predicts a compact Zn$_3$ cluster as the dominant configuration.
This result is consistent with recent low-temperature scanning tunneling microscopy (STM) measurements, in which supported Zn$_3$ species predominantly adopt triangular geometries.\cite{Rodriguez2025Nature}
Whether this structural preference persists under reaction conditions, however, remains unclear.

\paragraph{Finite-Temperature Reorganization of Zn$_3$(OH)$_3$ from Triangle to Line.}
To obtain an initial picture of the finite-temperature behavior, we first performed unbiased molecular dynamics simulations at 300 and 450 K.
All trajectories were initialized from the triangular configuration, with five independent 20 ns simulations performed at each temperature using random initial velocities sampled from the Maxwell--Boltzmann distribution.
The maximum Zn-Zn-Zn angle, $\theta_{\max}$, was used as an order parameter to distinguish the compact triangular ($\theta_{\max}\approx 60^\circ$) and extended linear ($\theta_{\max}\approx 170^\circ$) configurations.
Representative trajectories are shown in \textbf{Figure~\ref{fig:zn3x}b}.
At 300 K, the cluster remains trapped in the triangular basin throughout the simulation, indicating that the zero-temperature minimum remains kinetically stable on the nanosecond timescale.
In contrast, at 450 K the cluster rapidly transforms into the linear configuration and only rarely returns to the triangular state, revealing a qualitative temperature-induced structural reorganization that cannot be inferred from zero-temperature energetics alone.
The representative trajectories shown in \textbf{Figure~\ref{fig:zn3x}b} are reproduced across five independent simulations (\textbf{Figure~S5a}).
Although one trajectory exhibits a rare triangular-to-linear transition at 300 K, such events remain infrequent on the 20 ns timescale, confirming that the triangular configuration is the dominant kinetic state under these conditions.

To quantify this transition, we reconstructed the free-energy landscape using the on-the-fly probability enhanced sampling (OPES) method at 150, 300, and 450 K.
Two collective variables (CVs) were employed to describe the structural transformation of Zn$_3$-based clusters.
The primary CV is the maximum Zn$\text{-}$Zn$\text{-}$Zn angle $\theta_{\max}$ (defined in \textbf{Equation~\ref{eq:theta_max}}), defined as a smooth maximum over the three internal angles formed by the Zn atoms.
A secondary CV $\tilde{d}_{\max}$ (defined in \textbf{Equation~\ref{eq:d_tilde}}) was introduced to promote the sampling of the linear configuration, which is characterized by a larger maximum Zn$\text{-}$O distance than the triangular configuration.
The effectiveness of the selected CVs is further demonstrated by the corresponding time evolutions of the CVs and two-dimensional free-energy surfaces for Zn$_3$(OH)$_3$ and Zn$_3$(OH)$_2$CHOO on the pristine Cu(111), single-Zn-site, and alloyed Cu$_3$Zn surfaces at 150, 300, and 450 K (\textbf{Figures~S6-S11}).
The corresponding one-dimensional free-energy profiles obtained by projection onto $\theta_{\max}$ are summarized in \textbf{Figure~S12}.
For every system, the OPES simulations sample reversible transitions between the triangular and linear configurations, while the corresponding two-dimensional free-energy surfaces exhibit two well-defined basins associated with these configurations.
Each OPES simulation was performed for at least 25 ns.
To assess convergence, cumulative reweighted estimates were evaluated at 5 ns intervals; shaded regions, error bars, and reported uncertainties denote one standard deviation across these successive estimates.
The corresponding numerical values of $\Delta F$, $\Delta E$, and $T\Delta S$ are reported in \textbf{Table~S2}.

The resulting free-energy profiles of Zn$_3$(OH)$_3$ on pristine Cu(111) surface projected onto $\theta_{\max}$ are shown in \textbf{Figure~\ref{fig:zn3x}c}.
At 150 K, the free-energy minimum coincides with the triangular configuration, closely reproducing the zero-temperature prediction.
As the temperature increases, the free-energy minimum progressively shifts toward larger $\theta_{\max}$ values corresponding to the linear configuration, while the free-energy barrier separating the two basins decreases substantially.
By 450 K, the linear configuration becomes thermodynamically competitive, consistent with the configurational transition directly observed in the unbiased MD simulations.
The free-energy differences and corresponding Boltzmann populations are summarized in \textbf{Figure~\ref{fig:zn3x}d}.
Here, the $\Delta F$ at 0 K is approximated by the DP total energy difference between the optimized triangular and linear minimum-energy structures, while the finite-temperature free energies are obtained from OPES reweighting.
The negative $\Delta F$ values indicate that the linear configuration is favored.
The Boltzmann population (defined in \textbf{Equation~\ref{eq:boltzmann_population}}) represents the equilibrium probability of observing a given configuration at temperature, as determined by the free energy difference between the competing configurations.
Whereas the zero-temperature approximation predicts an exclusively triangular structure, finite-temperature sampling progressively stabilizes the linear ensemble.
At 150 K the triangular configuration remains dominant, while at 300 K the free-energy difference is already substantially reduced.
At 450 K, the free-energy difference approaches zero, giving rise to comparable equilibrium populations of the two configurations.
This comparison demonstrates that the configurational ensemble under operating conditions differs fundamentally from that predicted solely from minimum-energy structures.

To understand the origin of this temperature dependence, we further analyzed the potential-energy distributions sampled within each configurational basin (\textbf{Figure~\ref{fig:zn3x}e}).
Within the estimated standard deviations, the average potential-energy difference between the triangular and linear ensembles shows no systematic temperature dependence and remains close to the zero-temperature value over the entire temperature range.
In contrast, the entropy extracted from the free-energy analysis consistently favors the linear configuration, whose larger entropy increasingly offsets its enthalpic disadvantage as temperature rises.
The observed free-energy crossover therefore originates primarily from entropy rather than changes in the underlying potential-energy surface.
These results demonstrate that the catalytically relevant Zn$_3$ species should be viewed as a finite-temperature configurational ensemble rather than a static minimum-energy structure.

\begin{figure}
    \includegraphics[width=1.00\textwidth]{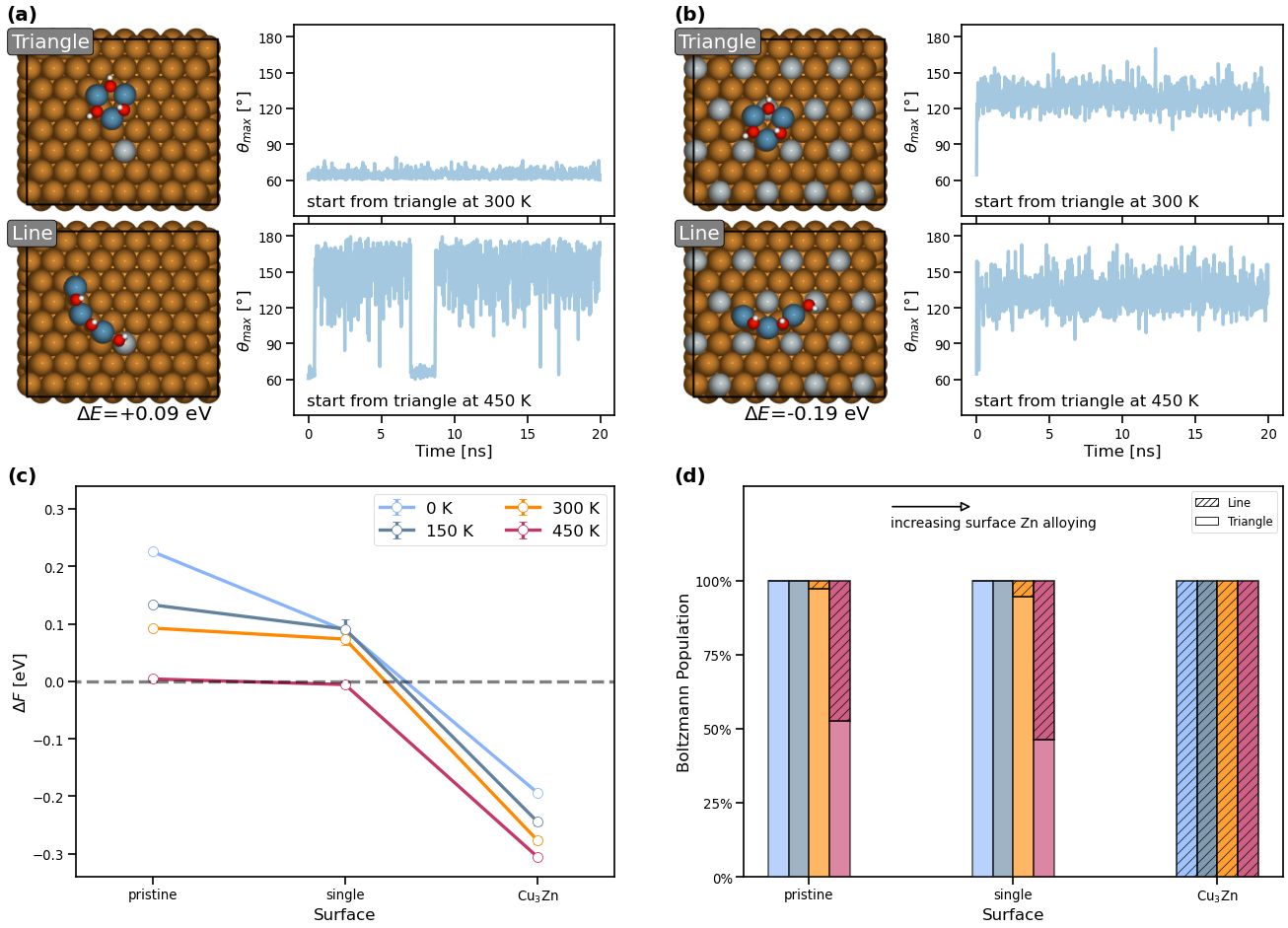}
    \caption{
        Effect of surface Zn alloying on the configurational stability of Zn$_3$(OH)$_3$.
        (a, b) Optimized triangular and linear configurations, together with representative unbiased MD trajectories at 300 and 450 K showing the evolution of the maximum Zn-Zn-Zn angle ($\theta_{\max}$), for the single-Zn-site surface (a) and the alloyed Cu$_3$Zn surface (b).
        The zero-temperature energy difference between the two configurations ($\Delta E = E_\mathrm{line} - E_\mathrm{triangle}$) is given below each structure. Atom colors in (a, b): Cu, brown; Zn in the cluster, blue; Zn in the surface, gray; O, red; H, white.
        (c) Free-energy difference between the linear and triangular configurations ($\Delta F = F_\mathrm{line}-F_\mathrm{triangle}$) as a function of surface composition at 0, 150, 300, and 450 K.
        Error bars denote one standard deviation across cumulative reweighted estimates evaluated at 5 ns intervals; where not visible, they are smaller than the symbols. No sampling uncertainty is assigned to the zero-temperature values.
        (d) Corresponding equilibrium Boltzmann populations of the triangular and linear configurations at 0, 150, 300, and 450 K as a function of surface composition.
    }
    \label{fig:alloy}
\end{figure}

\clearpage
\paragraph{Surface Alloying Effect on Configurational Stability.}
Having established the entropy-driven reorganization of Zn$_3$(OH)$_3$ on pristine Cu(111), we next investigate how surface Zn alloying modifies the configurational stability.
Representative 20 ns unbiased MD trajectories together with the corresponding 0 K minimum-energy structures are shown in \textbf{Figure~\ref{fig:alloy}a,b}, while all five independent trajectories are provided in \textbf{Figure~S5b,c}.

On the single-Zn-site surface (\textbf{Figure~\ref{fig:alloy}a}), the triangular configuration remains the zero-temperature minimum, although the energy difference relative to the linear configuration decreases from 0.23 eV on pristine Cu(111) to only 0.09 eV.
Unbiased MD simulations are consistent with this energetic ordering.
At 300 K, the cluster remains trapped in the triangular configuration throughout the 20 ns simulation, whereas at 450 K the linear configuration becomes dominant, exhibiting only occasional returns to the triangular state.
Thus, introducing an isolated Zn atom weakens the energetic preference for the compact triangular structure but does not qualitatively alter the finite-temperature behavior established on the pristine surface.
A different behavior is observed on the alloyed Cu$_3$Zn surface (\textbf{Figure~\ref{fig:alloy}b}).
Here, the energetic ordering is completely reversed, with the linear configuration becoming more stable than the triangular configuration by 0.19 eV at 0 K.
Consistent with this prediction, unbiased MD simulations show that the linear configuration is stabilized at both 300 and 450 K, with no distinct configurational transitions observed over the simulation timescale.
Compared with the pristine and single-Zn-site surfaces, the alloyed surface also exhibits substantially smaller angular fluctuations, indicating that extensive Zn alloying both stabilizes and rigidifies the linear Zn$_3$ configuration.

The free-energy differences and equilibrium Boltzmann populations of Zn$_3$(OH)$_3$ on three surfaces are summarized in \textbf{Figure~\ref{fig:alloy}c,d} with the OPES free-energy profiles of Zn$_3$(OH)$_3$ on pristine and single-Zn-site surfaces shown in \textbf{Figure~S12b,c}.
For both the pristine and single-Zn-site surfaces, increasing temperature progressively reduces the free-energy difference, leading to comparable populations of the triangular and linear configurations at 450 K.
In contrast, the alloyed Cu$_3$Zn surface maintains an overwhelming preference for the linear configuration over the entire temperature range.
The corresponding potential-energy distributions and the entropic contributions for the single-Zn-site and alloyed Cu$_3$Zn surfaces are shown in \textbf{Figure~S13b,c}.
Similar to the pristine surface, the average potential-energy difference between the two configurations exhibits only weak temperature dependence and remains close to the corresponding zero-temperature energy difference, confirming that the finite-temperature evolution of the free-energy landscape is governed primarily by entropy.

The entropic effect is also important for understanding the similar free-energy behavior on the pristine and single-Zn-site surfaces.
Although the linear structure on the single-Zn-site surface is additionally stabilized by formation of a Zn$_c$-OH-Zn$_s$ motif (Zn$_c$ denotes the cluster Zn atom and Zn$_s$ denotes the surface Zn atom), its free-energy profile differs only slightly from that on pristine Cu(111) surface.
To understand this behavior, we decomposed the linear ensemble into Zn-bound and Cu-bound configurations according to the metal-oxygen coordination environment and analyzed their reweighted populations (\textbf{Table~S3}).
Although Zn-bound linear structures are energetically more stable, the linear ensemble is dominated by Cu-bound configurations because the adsorbate can occupy many equivalent Cu sites but only a single Zn site.
The resulting entropy largely compensates for the enthalpic stabilization of the Zn-bound structure, producing a free-energy landscape similar to that of pristine Cu(111) surface.

Overall, these results demonstrate that increasing surface Zn alloying continuously reshapes the free-energy landscape of supported Zn$_3$ clusters.
While finite temperature universally promotes the linear configuration through entropy, extensive Zn alloying additionally stabilizes the linear structure energetically, making it the dominant configuration across the entire temperature range.

\begin{figure}
    \includegraphics[width=1.00\textwidth]{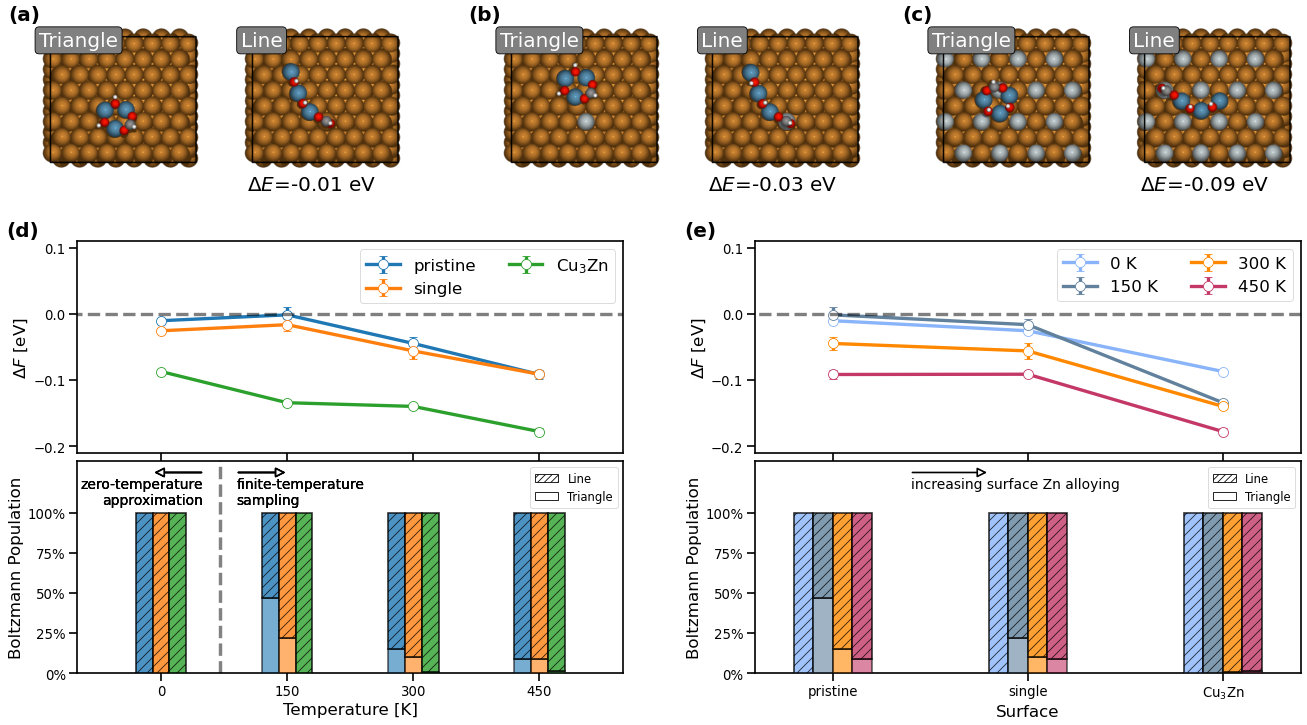}
    \caption{
        Temperature- and surface-dependent stability of Zn$_3$(OH)$_2$CHOO on Cu-based surfaces.
        (a–c) Optimized triangular and linear configurations on pristine Cu(111), the single-Zn-site surface, and the alloyed Cu$_3$Zn surface, respectively. The zero-temperature energy difference between the two configurations, $\Delta E = E_\mathrm{line}-E_\mathrm{triangle}$, is shown below each pair of structures. Atom colors in (a–c): Cu, brown; Zn in the cluster, blue; Zn in the surface, gray; O, red; C, black; H, white.
        (d) Free-energy difference between the linear and triangular configurations ($\Delta F = F_\mathrm{line}-F_\mathrm{triangle}$) as a function of temperature (top), together with the corresponding equilibrium Boltzmann populations (bottom).
        (e) Free-energy difference between the linear and triangular configurations as a function of surface composition at 0, 150, 300, and 450 K (top), together with the corresponding equilibrium Boltzmann populations (bottom).
        Error bars in the upper plots of (d) and (e) denote one standard deviation across cumulative reweighted estimates evaluated at 5 ns intervals; where not visible, they are smaller than the symbols. No sampling uncertainty is assigned to the zero-temperature values.
    }
    \label{fig:adsorbate}
\end{figure}

\clearpage
\paragraph{Adsorbate Effect on Configurational Stability.}
We next examine the influence of the reaction intermediate by considering Zn$_3$(OH)$_2$CHOO on the same three Cu-based surfaces.
The optimized triangular and linear configurations are shown in \textbf{Figures~\ref{fig:adsorbate}a-c}.
Compared with Zn$_3$(OH)$_3$, the presence of the formate ligand substantially reduces the energetic distinction between the two configurations.
The linear configuration is already slightly favored on the pristine Cu(111) and single-Zn-site surfaces by 0.01 and 0.03 eV, respectively, while alloying further increases its stabilization to 0.09 eV on the Cu$_3$Zn surface.
Thus, although the adsorbate shifts the energetic balance toward the linear configuration, it preserves the same qualitative alloying trend established for Zn$_3$(OH)$_3$.

The finite-temperature behavior of Zn$_3$(OH)$_2$CHOO was evaluated using the same unbiased MD and OPES protocols described above, and the complete set of $\theta_{\max}$ trajectories from unbiased MD simulations of the cluster on three surfaces is provided in \textbf{Figure~S5d-f}.
The resulting free-energy differences and Boltzmann populations are summarized from complementary temperature- and surface-dependent perspectives in \textbf{Figures~\ref{fig:adsorbate}d,e}, while the OPES free-energy profiles of Zn$_3$(OH)$_2$CHOO on three surfaces are shown in \textbf{Figure~S12d-f}.
The temperature and surface dependence closely mirrors that established for Zn$_3$(OH)$_3$.
Increasing temperature progressively stabilizes the linear configuration on all three surfaces, whereas increasing surface Zn content shifts the configurational equilibrium further toward the linear configuration at every temperature.
We further analyzed the corresponding potential-energy distributions and entropic contributions of Zn$_3$(OH)$_2$CHOO on three surfaces (\textbf{Figure~S13d-f}).
Similar to Zn$_3$(OH)$_3$, the average potential-energy difference between the triangular and linear configurations remains close to the corresponding zero-temperature value and exhibits only weak temperature dependence.
The finite-temperature evolution of the free-energy landscape is therefore again governed primarily by entropy rather than energetic changes.
The free-energy profiles on the pristine and single-Zn-site surfaces also remain similar despite the additional stabilization of the Zn$_\mathrm{c}$-CHOO-Zn$_\mathrm{s}$ motif on the single-Zn-site surface.
Similar to Zn$_3$(OH)$_3$, decomposition of the linear ensemble into Zn-bound and Cu-bound configurations (\textbf{Table~S3}) shows that Cu-bound structures dominate the sampled linear ensemble because the formate intermediate can occupy many equivalent Cu sites but only a single Zn site.
The resulting entropy compensates for the enthalpic stabilization of the Zn-bound configuration, producing free-energy landscapes that closely resemble those on pristine Cu(111) surface.

Consistent with the free-energy analysis, the linear configuration dominates the Boltzmann ensemble on all three surfaces at finite temperature, with only a small fraction of triangular structures remaining on the pristine and single-Zn-site surfaces.
The alloyed Cu$_3$Zn surface maintains an almost exclusively linear ensemble throughout the investigated temperature range.
Overall, Zn$_3$(OH)$_2$CHOO follows the same temperature- and surface-dependent evolution established for Zn$_3$(OH)$_3$, while the presence of the formate intermediate further shifts the configurational equilibrium toward the linear ensemble.

\clearpage
\paragraph{Spatial Stability of Zn$\mathbf{_3}$ Clusters against Sintering.}
Having established the configurational stability of supported Zn$_3$ clusters through finite-temperature free-energy landscapes, we next examine their spatial stability on the catalyst surface.
Configurational stability determines which configuration are dominantly populated at equilibrium, whereas spatial stability describes the ability of these active sites to remain spatially isolated under reaction conditions.\cite{Dai2018ChemSocRev}
Even thermodynamically stable Zn$_3$ clusters may become catalytically ineffective if they diffuse across the surface, collide with neighboring clusters, and undergo coalescence.
Therefore, characterizing the finite-temperature mobility of supported Zn$_3$ clusters provides direct insight into their resistance to sintering and the long-term stability of the active-site ensemble.

\begin{figure}
    \includegraphics[width=1.00\textwidth]{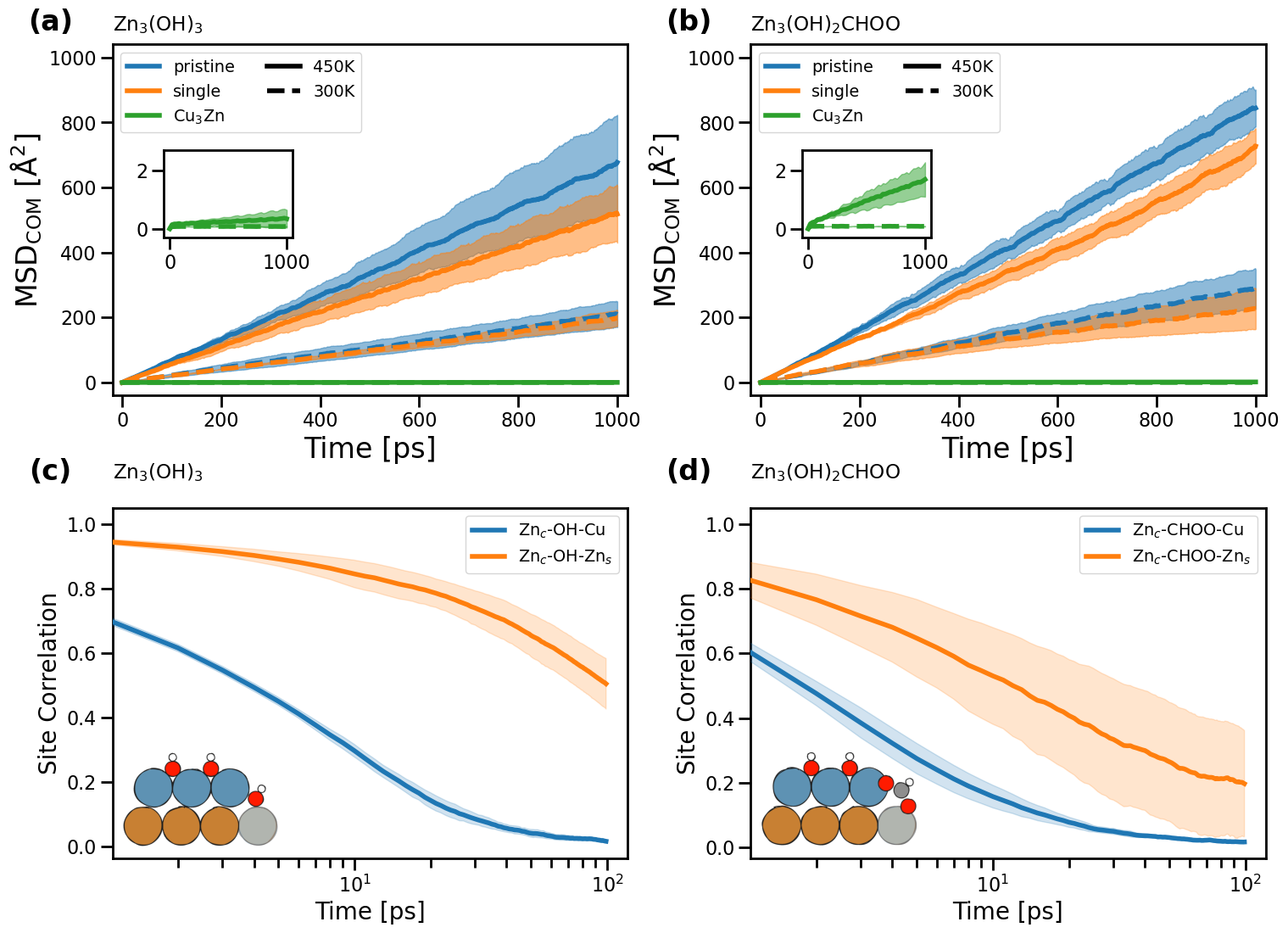}
    \caption{
        Mobility analysis of Zn$_3$ clusters on different surfaces.
        (a,b) Mean squared displacement (MSD) of the cluster center of mass (COM) as a function of time for Zn$_3$(OH)$_3$ (a) and Zn$_3$(OH)$_2$CHOO (b) on pristine Cu(111), the single-Zn-site surface, and the alloyed Cu$_3$Zn surface at 300 K (dashed) and 450 K (solid).
        Insets show the MSD on the alloyed Cu$_3$Zn surface on an enlarged scale.
        Shaded regions represent the standard deviation from five independent trajectories.
        (c,d) Site correlation functions for Zn$_3$(OH)$_3$ (c) and Zn$_3$(OH)$_2$CHOO (d) on the single-Zn-site surface at 450 K.
        The site correlation function gives the probability that the cluster remains bound to the same surface site as a function of time.
        Results are shown separately for Zn$_c\text{-}$OH$\text{-}$Cu and Zn$_c\text{-}$OH$\text{-}$Zn$_s$ bridge sites in (c), and Zn$_c\text{-}$CHOO$\text{-}$Cu and Zn$_c\text{-}$CHOO$\text{-}$Zn$_s$ bridge sites in (d), where Zn$_c$ denotes a Zn atom in the cluster and Zn$_s$ a Zn atom in the surface.
        Shaded regions represent the standard deviation from five independent trajectories.
        The schematic insets in (c,d) illustrate the bridge-site geometries used to define the site correlation functions. Blue, brown, gray, black, red, and white spheres in (c,d) denote Zn$_c$, Cu, Zn$_s$, C, O, and H, respectively.
        The rightmost surface atom can be either Cu or Zn$_s$, defining the corresponding bridge sites.
    }
    \label{fig:dynamic_mobility}
\end{figure}

To quantify spatial stability, we analyzed unbiased MD trajectories using the same simulations employed above for the angular evolution.
For both Zn$_3$(OH)$_3$ (\textbf{Figure~\ref{fig:dynamic_mobility}a}) and Zn$_3$(OH)$_2$CHOO (\textbf{Figure~\ref{fig:dynamic_mobility}b}), the mean squared displacement of the cluster center of mass (MSD$_\mathrm{COM}$), defined in \textbf{Equation~\ref{eq:msd_com}}, increases approximately linearly with time on the pristine Cu(111) and single-Zn-site surfaces, indicating continuous lateral diffusion of the supported clusters.
As expected, increasing the temperature from 300 K to 450 K significantly enhances mobility, reflected by the steeper MSD$_\mathrm{COM}$ slopes.
In contrast, diffusion on the alloyed Cu$_3$Zn surface is strongly suppressed, with negligible MSD$_\mathrm{COM}$ over the entire simulation timescale, suggesting that the higher Zn concentration stabilizes the clusters and restricts their lateral motion.
This trend is consistent across both cluster compositions, although Zn$_3$(OH)$_2$CHOO exhibits slightly increased mobility compared to Zn$_3$(OH)$_3$.
The above results highlight the critical role of alloying in suppressing cluster mobility and demonstrate that both thermal effects and surface composition must be considered when assessing the dynamic stability of active sites in the CZA catalyst under reaction conditions.

To elucidate the microscopic origin of the reduced mobility on single-Zn-site surface, we analyzed the site correlation functions of the supported clusters at 450 K.
The site correlation (defined in \textbf{Equation~\ref{eq:site_correlation}}) measures the probability that an adsorbate remains bound to the same surface site over time, providing a direct probe of residence stability.
Here, we distinguish two types of interfacial bridging motifs: Zn$_c\text{-}$adsorbate$\text{-}$Zn$_s$ and Zn$_c\text{-}$adsorbate$\text{-}$Cu, where Zn$_c$ denotes a Zn atom within the cluster and Zn$_s$ denotes a Zn atom in the surface layer.
At 300 K, Zn$_3$(OH)$_3$ remains predominantly in the triangular configuration on both the pristine Cu(111) and single-Zn-site surfaces, and thus does not form well-defined Zn$_c\text{-}$OH-Cu or Zn$_c\text{-}$OH-Zn$_s$ bridging motifs. 
In contrast, Zn$_3$(OH)$_2$CHOO undergoes frequent transitions between triangular and linear configurations, allowing both bridging motifs to form already at 300 K.
However, for a consistent and quantitative comparison between the two supported clusters, it is necessary to consider conditions where both bridging motifs are simultaneously well-defined and dynamically relevant.
Therefore, we focus on the 450 K regime on the single-Zn-site surface, where the linear configuration is predominantly stabilized and both Zn$_c\text{-}$adsorbate$\text{-}$Cu and Zn$_c\text{-}$adsorbate$\text{-}$Zn$_s$ motifs coexist.
Under these conditions, the site correlation analysis reveals a clear distinction between the two binding environments.
For both Zn$_3$(OH)$_3$ (\textbf{Figure~\ref{fig:dynamic_mobility}c}) and Zn$_3$(OH)$_2$CHOO (\textbf{Figure~\ref{fig:dynamic_mobility}d}), the correlation associated with Zn$_c\text{-}$Cu bridge sites decays rapidly, whereas the Zn$_c\text{-}$Zn$_s$ correlation exhibits a much slower decay.
This contrast indicates that adsorption on Zn$_c\text{-}$Cu sites is comparatively labile, with frequent site-to-site hopping, while Zn$_c\text{-}$Zn$_s$ bridge sites serve as more stable anchoring centers that retain the cluster over longer timescales.
These findings are consistent with the MSD analysis, where enhanced lateral diffusion arises from repeated detachment from weaker Zn$_c\text{-}$Cu sites and reattachment at neighboring sites.

Interestingly, this kinetic stabilization is largely decoupled from configurational stability.
As shown above, the single-Zn-site surface exhibits free-energy landscapes similar to those of pristine Cu(111), indicating only a modest effect on the equilibrium populations of triangular and linear configurations.
Nevertheless, the presence of an isolated surface Zn atom prolongs the lifetime of the Zn$_c\text{-}$adsorbate$\text{-}$Zn$_s$ bridge and suppresses cluster migration.
This kinetic stabilization may help prevent sintering and maintain catalytically relevant Zn$_3$ ensembles under reaction conditions.

\newpage
\section{Conclusion}
The configurational stability of supported Zn$_3$ clusters is governed by competing enthalpic and entropic effects.
Compact triangular structures are generally favored by the 0 K energetics, whereas increasing temperature stabilizes extended linear ensembles through their greater entropy.
Formate coordination and extensive Zn alloying of the Cu surface shift this balance further toward the linear configuration.
At dilute Zn content, configurational entropy offsets the enthalpic preference for Zn-bound motifs, producing a free-energy landscape similar to pristine Cu(111).
Thus, structures that appear unfavorable in static calculations can become thermodynamically relevant under reaction conditions.
The surface also controls cluster mobility independently of the configurational equilibrium.
On pristine Cu(111), Zn$_3$ species remain mobile and may migrate and coalesce, whereas surface Zn forms longer-lived Zn$_\mathrm{c}$--adsorbate--Zn$_\mathrm{s}$ anchoring motifs and strongly suppresses lateral diffusion.
The Cu--Zn interfacial composition therefore both reshapes the free-energy landscape and helps preserve isolated Zn species against sintering.

These findings connect with experiments on CZA catalysts for methanol-synthesis.
The low-temperature preference for triangular Zn$_3$ structures is consistent with recent cryogenic STM observations,\cite{Rodriguez2025Nature} while the predicted emergence of extended and dynamic ensembles near operating temperatures provides a target for operando characterization.
Although Zn$_3$(OH)$_3$ and Zn$_3$(OH)$_2$CHOO on idealized Cu-based surfaces do not capture the full complexity of working CZA catalysts, they isolate how temperature, adsorbates, and local Cu--Zn composition jointly determine the structure and persistence of interfacial Zn species.
These trends can guide experimental interpretation and the construction of more realistic CZA active-site models, without assigning catalytic activity to a particular Zn$_3$ configuration.

More broadly, static minimum-energy structures alone may misrepresent active-site configurations.
Although quantitative energetics and crossover temperatures may depend on the exchange--correlation functional, finite-temperature entropy can reorganize active-site ensembles in ways that static optimization cannot predict.
Catalytic sites should therefore be viewed as dynamic ensembles whose populations and surface mobility evolve with temperature, adsorbate coverage, and local composition.
MLIP-accelerated free-energy sampling and long-timescale MD provide a practical route to capturing this behavior, thereby bridging idealized static models and the complex catalyst structures present under operating conditions.

\newpage
\section{Methods}
\paragraph{Density Functional Theory.}
All DFT calculations are performed with the Vienna Ab initio Simulation Package (VASP),\cite{Kresse1996ComputMaterSci,Kresse1996PRB} using PBE approximation for the exchange-correlation functional.\cite{Perdew1996PRL}
The projector augmented wave method (PAW)-PBE\cite{Kresse1999PRB} is adopted to describe the core-valence interaction of the electrons.
To sample the Brilluoin zone, we use $\Gamma$-centered uniform $k$-point meshes with a spacing of 0.04 \AA$^{-1}$ in each direction of the reciprocal space.\cite{Monkhorst1976PRB}
To account for van der Waals interactions, we use the correction of D3 with zero damping.\cite{Grimme2010JCP,Grimme2011JComputChem}
The D3 correction on top of PBE is computed with the simple-dftd3 package.\cite{Ehlert2026SimpleDFTD3}

\paragraph{Machine Learning Interatomic Potential.}
The MLIP is trained with the DP architecture\cite{Zhang2018PRL,Zhang2018NeurIPS} implemented in the DeePMD-kit package.\cite{Wang2018CPC,Zeng2023JCP}
The DP architecture employs neural networks to map the local atomic environment to the atomic energy.
The local atomic environment is defined by neighboring atoms within a radius cutoff of 6 \AA.
The ``se\_e2\_a'' descriptor is adopted to represent the local atomic environment with an embedding network of $24 \times 48 \times 96$.
The dimension of the axis neuron is set to 16.
The fitting network from the descriptor to the atomic energy is $256 \times 256 \times 256$.

The training dataset spans three categories: (i) pure metals, including Cu bulks (Cu$_{16}$, Cu$_{32}$) and surfaces (Cu$_{12}$, Cu$_{16}$), Zn bulks (Zn$_{48}$, Zn$_{64}$), as well as CuZn alloy bulks (Cu$_{8}$Zn$_{8}$, Cu$_{24}$Zn$_{8}$) and surfaces (Cu$_{48}$Zn$_{48}$, Cu$_{48}$Zn$_{16}$); (ii) metal oxides, represented by bulk ZnO (Zn$_{48}$O$_{48}$); and (iii) supported clusters, including ZnO$_\mathrm{x}$, Zn(OH)$_\mathrm{y}$, and Zn(OH)$_\mathrm{y}$(CHOO)$_\mathrm{z}$ on Cu and CuZn surfaces.
Although the present work focuses on Zn$_3$ clusters, the DP model is constructed with a broader and transferable scope.
Accordingly, the dataset includes configurations with cluster sizes up to Zn$_{17}$ and diverse surface environments, enabling the model to robustly describe cluster-surface and cluster-adsorbate interactions beyond the specific clusters examined here.
For supported clusters, several four-layered (111) slabs, $p(4 \times \sqrt{3})$, $p(6 \times 3\sqrt{3})$ and $p(8 \times 4\sqrt{3}))$, are employed to accommodate clusters of different sizes and spatial extents, capturing cluster size and surface alloying effects.
The final dataset contains 44,269 structures.
The model reaches root-mean-squared errors of 0.005 eV/atom for energies and 0.056 eV/Å for forces.

\paragraph{Local and Global Optimization.}
All structural optimization tasks are performed with GDPy\cite{Xu2026GDPy}, which interfaces both DP and VASP calculations with the BFGS\cite{Bonnans2006NumericalOptimization} algorithm implemented in ASE.\cite{Larsen2017JPhysCondensMatter}
The global optimization are performed with the genetic algorithm implemented in GDPy.
The structures in the initial generation are randomly generated while the ones in the following generations are constructed by crossover and mutation.
The crossover operation generates a new structure by a combination of two parent structures by cut-and-splice.\cite{Deaven1995PRL}
The mutation operation is performed on the offspring structures by applying with equal probability among three types of mutation: (1) random displacement of atoms, (2) random translate of the cluster, and (3) random rotate of the cluster.
For each system, we employ DP to run 20 generations with a population size of 50, resulting in a total of 1000 structures.
The key configurations are further optimized with DFT to validate the energetic ordering.

\paragraph{Molecular Dynamics and Enhanced Sampling.}
All MD simulations are performed with LAMMPS.\cite{Thompson2022ComputPhysCommun}
The timestep is set to 0.5 fs.
The NVT ensemble is adopted with a Nos\'{e}-Hoover chain\cite{Martyna1992JCP} thermostat with a damping parameter of 100 fs.
These settings are applied to both unbiased MD and enhanced sampling simulations listed below.

To compute the free energy differences between different configurations at finite temperatures, we employed the OPES method\cite{Invernizzi2020JPCL} implemented in PLUMED.\cite{Tribello2014CPC}
Two CVs are used to describe the structural transformation between the triangular and linear configurations of the Zn$_3$ clusters.
The first CV is the maximum Zn–Zn–Zn angle ($\theta_\mathrm{max}$), which is defined as
\begin{equation}\label{eq:theta_max}
    \theta_\mathrm{max} = \beta \log \sum_{i=1}^3 \exp (\theta_i / \beta)
\end{equation}
where $\theta_i$ is the angle formed by the three Zn atoms in the cluster, and $\beta$ is a parameter set to 50 in the log-sum-exp function to ensure a smooth maximum.
This CV provides a continuous measure of the cluster geometry, distinguishing compact triangular configurations from extended linear configurations, and is used as the order parameter for free-energy projections.
The second CV is a distance-based descriptor ($\tilde{d}_\mathrm{max}$) defined as the smooth maximum of the Zn$\text{-}$O distances, which is defined as
\begin{equation}\label{eq:d_tilde}
    \tilde{d}_{max} = \beta \log \sum_{i=1}^3 \exp(d_{max}^i / \beta)
\end{equation}
where $d_\mathrm{max}^i$ is the maximum distance between the $i$-th O-contained species (hydroxyl or formate) and the three Zn atoms in the cluster, and $\beta$ is set to 50 as well.
By promoting fluctuations along bond-length degrees of freedom, this auxiliary CV accelerates transitions between the triangular and linear configurations.
However, it does not directly define the structural states of interest and is therefore not used in the final free-energy analysis.
The quantities $d_\mathrm{max}^i$ and $d_\mathrm{min}^i$ are defined as smooth maximum and minimum distances, respectively:
\begin{equation}
    d_\mathrm{max}^i = \beta \log \sum_{j=1}^2 \exp((||R_i-R_j||) / \beta)
\end{equation}
\begin{equation}
    d_\mathrm{min}^i = \beta / (\log \sum_{j=1}^2 \exp(\beta / (||R_i-R_j||)))
\end{equation}
where $R_i$ and $R_j$ are the positions of the $i$-th O-contained species and the $j$-th Zn atom of the two Zn atoms bonded to the O-contained species in the initial triangular configuration, respectively.
For hydroxyl, $R_i$ is the position of the O atom, while for formate, $R_i$ is the average position of the two O atoms in the formate.
The smoothing parameter $\beta$ is set to 50 and 100 for $d_\mathrm{max}^i$ and $d_\mathrm{min}^i$, respectively, to ensure a smooth maximum and minimum.
In addition, to prevent undesirable dissociation events e.g. Zn$_3$(OH)$_3$ $\rightarrow$ Zn$_3$(OH)$_2$ + OH, harmonic upper-wall restraints are applied to both $d_\mathrm{max}^i$ and $d_\mathrm{min}^i$ with force constants of 1000 kJ/mol/\AA$^2$ and a distance cutoff of 10.0 \AA{} and 3.0 \AA, respectively, effectively confining the sampling to chemically relevant configurations.
The OPES bias potential is constructed adaptively with a target barrier height of 30 kJ/mol and deposited every 500 steps.
The initial Gaussian widths are chosen based on the fluctuations of the CVs from the initial unbiased 40,000 steps.

\paragraph{Population Analysis.}
The Boltzmann population of each configuration is computed from the free energy difference between the linear and triangular configurations.
For a many-state system, the probability of observing configuration i at temperature T is given by
\begin{equation}\label{eq:boltzmann_population}
    P_i(T) = \frac{\exp(-\Delta F_i / k_B T)}{\sum_j \exp(-\Delta F_j / k_B T)}
\end{equation}
where $\Delta F_i$ is the free energy of configuration i relative to the reference state, $k_B$ is the Boltzmann constant, and $T$ is the temperature.
Here, the triangular configuration is taken as the reference ($\Delta F=0$).
At 0 K, the free-energy difference is approximated by the total energy difference between the linear and triangular configurations.
The corresponding Boltzmann population is evaluated at 5 K as a numerical approximation to the zero-temperature limit, thereby avoiding divergence in the Boltzmann factor.

\paragraph{Mobility Analysis.}
The cluster mobility is measure by the mean square displacement (MSD) of the center of mass (COM) of the cluster, which is defined as
\begin{equation}\label{eq:msd_com}
    \mathrm{MSD}_\mathrm{COM}(t; R) = \langle\frac{1}{N} \sum_{i=1}^N \|R(t+\tau) - R(\tau)\|^2\rangle_{\tau}
\end{equation}
where $N$ is the number of clusters in the system, $R(t)$ is the position of the COM of the cluster at time $t$, and $\langle...\rangle_\tau$ denotes the average over all time origins $\tau$ with a lag time of 0.1 ns.
In all simulations, the number of clusters $N$ is 1.

To analyze the site stability and hopping dynamics of the clusters, we use the site correlation functions $C(t)$ defined by
\begin{equation}\label{eq:site_correlation}
    C(t) = \langle\frac{\langle h_{ij}(t+\tau)h_{ij}(\tau)\rangle_{ij}}{\langle h_{ij}(\tau)h_{ij}(\tau)\rangle_{ij}}\rangle_{\tau},
\end{equation}
where $h_{ij}=1$ if an adsorbate $i$ stays on the site $j$ and $h_{ij}=0$ otherwise, $\langle\dotsb\rangle_{ij}$ denotes the average over all the adsorbate-site present when $t=\tau$, the time origin, and $\langle\dotsb\rangle_{\tau}$ denotes the average over all time origins with a lag time of 1 ps.
To evaluate the site correlation, a criterion is required to determine whether an adsorbate is bound to a given bridge site.
In this work, a bridge binding configuration is defined based on geometric criteria involving the coordination between the adsorbate oxygen atoms and metal atoms.
Here, Zn$_c$ denotes a Zn atom within the cluster, while Zn$_s$ denotes a Zn atom in the surface.
For OH, a Zn$_c\text{-}$Cu or Zn$_c\text{-}$Zn$_s$ bridge is considered to be formed when the O atom simultaneously coordinates to both metal atoms, i.e., when both the Zn$_c\text{-}$O and Cu-O (or Zn$_s\text{-}$O) distances are smaller than 2.4 Å.
For CHOO, the bridge binding configuration involves either a Zn$_c\text{-}$O$\text{-}$CH$\text{-}$O$\text{-}$Cu or Zn$_c\text{-}$O$\text{-}$CH$\text{-}$O$\text{-}$Zn$_s$ motif.
In this case, the two oxygen atoms of the formate group play distinct roles: one O atom must coordinate to Zn$_c$, while the other O atom coordinates to either Cu or Zn$_s$.
A bridge site is therefore identified when the Zn$_c\text{-}$O distance for one oxygen and the Cu$\text{-}$O (or Zn$_s\text{-}$O) distance for the other oxygen are both smaller than 2.4 Å.

\newpage
\begin{acknowledgement}
The funding for this work was provided by Shell International Exploration and Production Inc., USA.
Most of the calculations were performed using resources provided by Princeton Research Computing at Princeton University and by the Computational Chemical Science Center ``Chemistry in Solution and at Interfaces'' (CSI), supported by the U.S. Department of Energy under Award No. DE-SC0019394.
The authors also acknowledge the National Energy Research Scientific Computing Center (NERSC) for additional computational resources provided through award ERCAP0021510 under Contract No. DE-AC02-05CH11231.
\end{acknowledgement}

\begin{suppinfo}
    Figures 1-13, Tables 1-3, and notes.
\end{suppinfo}

\section{Competing Interests}
The authors declare no competing financial or non-financial interests.

\clearpage
\bibliography{references}

\end{document}